 \documentclass[final,3p,times,twocolumn]{elsarticle}

\usepackage{graphicx}
\usepackage{epsfig}
\usepackage{hyperref}
\usepackage{amssymb}

\journal{Computer Physics Communications}

\begin{document}

\begin{frontmatter}



\title{Modelling opinion formation driven communities in social networks}


\author[BECS]{Gerardo I\~{n}iguez}
\author[UNAM,BECS]{Rafael A. Barrio}
\author[BUTE,BECS]{J\'anos Kert\'esz}
\author[BECS]{Kimmo K. Kaski\corref{kk}}
\ead{kimmo.kaski@hut.fi}

\address[BECS]{Centre of Excellence in Computational Complex Systems Research, Department of Biomedical Engineering and Computational Science, Aalto University School of Science and Technology, Espoo, Finland}
\address[UNAM]{Instituto de F\'{\i}sica, Universidad Nacional Aut\'onoma de M\'exico, Apartado Postal 20-364, 01000 M\'exico, Distrito Federal, Mexico }
\address[BUTE]{Institute of Physics, Budapest University of Technology and Economics, Budapest, Hungary}
\cortext[kk]{corresponding author}
\begin{abstract}
In a previous paper we proposed a model to study the dynamics of opinion formation in human societies by a co-evolution process involving two distinct time scales of fast transaction and slower network evolution dynamics. In the transaction dynamics we take into account short range interactions as discussions between individuals and long range interactions to describe the attitude to the overall mood of society. The latter is handled by a uniformly distributed parameter $\alpha$, assigned randomly to each individual, as quenched personal bias. The network evolution dynamics is realized by rewiring the societal network due to state variable changes as a result of transaction dynamics. The main consequence of this complex dynamics is that communities emerge in the social network for a range of values in the ratio between time scales. In this paper we focus our attention on the attitude parameter $\alpha$ and its influence on the conformation of opinion and the size of the resulting communities. We present numerical studies and extract interesting features of the model that can be interpreted in terms of social behaviour. 
\end{abstract}


\end{frontmatter}


\section{Introduction}
\label{sec:1}
Over the last decade various complex systems, from biology to human societies, have 
been approached from the perspective of network theory, with significant contributions 
to our understanding of their structure, function, and response~\cite{newman2006sad,boccaletti2006cns,caldarelli2007sfn,dorogovtsev2002enf}. 
In the case of human social systems, this approach was first adopted by social scientists~\cite{wasserman1994sna,baron2010sps,granovetter1985eas} 
studying relatively small-scale data sets collected from e.g. questionnaires 
and then by physicists~\cite{watts1998cds,albert1999dww} focusing on large-scale data sets 
collected from Internet, emails etc. Both these approaches give insight into social network structure with 
one-to-one interactions. In social sciences the social network paradigm views {\it social life as consisting 
of the flow and exchange of norms, values, ideas, and other social and cultural resources channeled 
through a network}~\cite{white1976ssm}. This obviates an answer to the question 'why large-scale 
social networks', being that studying collective social phenomena such as diffusion and spreading processes 
(epidemics)~\cite{pastorsatorras2001ess}, opinion formation~\cite{holley1975etw,araripe2009rpv}, and evolution of language~\cite{abrams2003mdl,castello2006odt} etc. is interesting and requires 
large systems. Thus we can ask (i) how are large scale social networks organised 
and (ii) whether they can be modelled with a simple model~\cite{castellano2009sps,sobkowicz2009mof}.
  
Everyday experience and empirical findings like those from mobile communication based social network~\cite{onnela2007sat} 
prove that society consists of communities of different sizes. Modeling the formation of such modules remains a focus issue in this context \cite{moreira2006ccg,kumpula2007prl}. Recently we posed the question: How does opinion dynamics influence and finally result in the community structure of a society?

We built a model~\cite{iniguez2009ocf}, in which 
individual-level microscopic opinion based mechanisms translate to forming on one hand mesoscopic communities 
and on the other hand the whole system at the macroscopic level. In this model we took basic notions from sociology 
to propose rules by which the social network changes in time, as {\it cyclic closure} forming ties with one's 
network neighbours, i.e. 'friends of friends',  and {\it focal closure} forming ties independently of the geodesic distance through 
shared activities, e.g. hobbies~\cite{kossinets2006eae}.  The main feature of our model is that we neatly separate the dynamics of the opinion changes of the individual agents from the changes in network connections. Then the appearance of distinguishable communities is driven by the dynamics taking place at two different time scales. A key ingredient to the model is the introduction of an ``attitude parameter" ($\alpha_i$) that mimics the behaviour of an individual $i$ facing an overall trend of opinion in the network. 

In this paper we focus our attention on the role of $\alpha_i$ that was found important in building communities sharing the same opinion~\cite{iniguez2009ocf}, and turned out to affect the relative size of communities. In what follows we succinctly explain the model and present numerical calculations.

\section{Model}
\label{model}
We consider opinion formation in a network of a fixed number of individuals or 
agents ($N$) to whom a simple question is posed. A state variable $x_i\in{[-1,1]}$ is associated with each individual $i$, which measures the agent's instantaneous inclination concerning the question at hand, while the elements $A_{ij}\in{\lbrace{1,0}\rbrace}$ of the network's connectivity (adjacency) matrix represent the presence or absence of discussions between individuals related to this question.

The time scale for discussions or exchange of information between individuals ( called ``transactions") is $dt$, while the time scale for a generalised change of connections in the network (called ``generation") is $T$. These two quantities are related by $T=gdt$, where the parameter $g$ defines the number of transactions per generation. These mutually dependent processes, or co-evolution, can be described in general with a dynamical equation of the state variable $x_i$ of agent $i$ as follows
\begin{equation}
\label{eq:1}
\frac{dx_i}{dt}=\frac{\partial x_i}{\partial t} + \sum_j \hat{O}(x_i,x_j,g) A_{ij},
\end{equation}
where the operator $\hat{O}$ acts discretely on the network only at time intervals $T=gdt$. Then we write for the first term of Eq.~\ref{eq:1}

\begin{equation}
\label{eq:microdyn}
\frac{\partial{x_i}}{\partial{t}} = \alpha_i f_0(\lbrace{x_j}\rbrace) + f_1(\lbrace{x_j}\rbrace) x_i,
\end{equation}
with
\begin{equation}
\label{eq:long}
f_0 = \sum_{\ell=2}^{\ell_{max}} \frac{1}{\ell} \sum_{j \in m_{\ell}(i)} x_j,
\end{equation}
and
\begin{equation}
\label{eq:short}
f_1 = \textrm{sgn}(x_i) \sum_{j \in m_1(i)} x_j,
\end{equation}
where $\alpha_i$ measures the attitude of agent $i$ towards overall or public opinion, $m_{\ell}(i)$ means the set of $\ell^{th}$ neighbors of $i$, and $\ell_{max}$ is the number of steps needed to reach the most distant neighbors of $i$.
\medskip

When $T\sim{dt}$, the dynamics of the state variables are irrelevant and we can concentrate on the discrete network evolution defined by a set of rules for deleting and creating links between agents. Agent $i$ can choose to cut an existing link with agent $j$, i.e. end a discussion if their opinions are incompatible. In order to perform this process the quantity
\begin{equation}
\label{eq:cut}
p_{ij} = A_{ij} \frac{|x_i - x_j|}{2}
\end{equation}
is calculated and all the links are put in ranking order. Then the links with larger $p_{ij}$-values are 
deleted, since they correspond to opinion divergence between the pair of agents. After the link deletion step follows the link creation step, where an agent $i$ can create a link with a second neighbour by starting a discussion 
with the ``friend of a friend" if this new link can help the agent in reaching a state of total conviction
($|x_i| = 1$). In order to determine this we calculate the quantity
\begin{equation}
\label{eq:create}
q_{ij} = \left(1 - A_{ij}\right) \Theta \left[(A^2)_{ij}\right] \frac{|x_i + x_j|}{2},
\end{equation}
where $\Theta[x] = \sum_{k=1}^{N-2} \delta_{k,x}$ and put them to ranking order. Then the links with larger $q_{ij}$-values are created. The link cutting and creation is performed by keeping their numbers the same, with a procedure explained in detail in Section 3.

The system is initialised to a random network configuration of $N$ nodes with an average degree $\left<k_0\right>$, and its evolution follows a two-step process. In the first step the dynamics of transactions is realised according to Eq.~(\ref{eq:microdyn}). By keeping the parameters fixed the system is driven until the specified time ($T=gdt$) or $g$ time steps to then do the second step, namely the network rewiring described above. This two-step process is iterated until the system reaches its final state, where no more changes in the state variables (all $x_i$'s) and in the network connections take place.

\section{Numerical procedure}
\label{num}
At a certain time $t$, the system is defined by a state vector $\mathbf{x}(t)=[x_1(t),x_2(t),...,x_i(t),...,x_N(t)]$ with entries $x_i(t)\in[-1,1]$ and a connectivity matrix $\mathbf{A}(t)$ with elements $A_{ij}(t)=0,1$, where $i,j=1,...,N$ and $N$ is the size of the network.

Each numerical process comprises $M+1$ steps where M is the smallest integer such that
\begin{equation}
\label{eq:StdDev}
\sum_{i=1}^{N}\left|x_i(M)-x_i(M-1)\right|<\Delta
\end{equation}
with $\Delta=10^{-6}$, and is essentially divided in four actions: (1) setting of initial conditions, (2) numerical integration of the dynamical equations, (3) keeping track of agents with irrevocable opinions ($\pm1$), and (4) rewiring of the network. Next we shall describe each one of them in more detail.

(1) For $t_0=0$, the components of $\mathbf{x}(t_0)$ are set randomly according to a normal probability distribution characterised by variance $v_0=1$ and mean $m_0=0$. Accordingly, the elements of $\mathbf{A}(t_0)$ are set randomly in such a fashion that $A_{ij}(t_0)=A_{ji}(t_0)$, $A_{ii}(t_0)=0$ and
\begin{equation}
\label{eq:IniAvDeg}
\frac{1}{N}Tr\mathbf{A}^2(t_0)=\langle k_0\rangle,
\end{equation}
where $\langle k_0\rangle$ is the initial average degree of the network. In all simulations
presented here we used $\langle k_0\rangle = 4$.

(2) For every $t+1$ such that $t=0,...,M-1$, we find the entries in $\mathbf{x}(t)$ that fulfil the condition $|x_i(t)|<1$ and solve their corresponding dynamical equations using a simple Euler method, that is to say, we set
\begin{equation}
\label{eq:SolDynEq1}
x_i(t+1)=x_i(t)+dt\left[\alpha_if_0(t)+f_1(t)x_i(t)\right],
\end{equation}
where $dt=10^{-4}$ is the time step (small enough to avoid artefacts). Each $\alpha_i\in[-1,1]$ is a random constant initially chosen from a uniform probability distribution, and the short and long range interaction terms are given by Eqs.~\ref{eq:long} and~\ref{eq:short}.

(3) In every time step we find the entries of $\mathbf{x}(t)$ that fulfil the condition $|x_i(t)|\geq1$ and set them to $x_i(t+1)=\textrm{sgn}(x_i(t)).$

(4) For every $t+1=ng<M+1$ where $g=1000$ is the generation time parameter and $n=1,2,...$, we find those entries of $\mathbf{x}(t+1)$ that fulfil the condition $|x_i(t+1)|<1$ and for each of them we construct the two sets $S_l=\left\lbrace{l_{ij}|l_{ij}>c}\right\rbrace$ with $l=p,q$ and a weight cut-off $c=0$, where the local rewiring weights $p_{ij},q_{ij}$ are given by
Eqs.~\ref{eq:cut}, and~\ref{eq:create}.

Then we get the minimum value between the cardinalities of the $S_p,S_q$ sets, namely $n_c=\min{\left\lbrace{S_p,S_q}\right\rbrace}$, and construct the subsets $C_l$ comprising the $n_c$ largest elements in the sets $S_l$. Denoting by $j'$ the second index in the elements $p_{ij'}\in{C_p}$ and by $j''$ the second index in the elements $q_{ij''}\in{C_q}$, we finally perform the rewiring operations over the connectivity matrix $\mathbf{A}(t+1)$, that is to say
\begin{equation}
\label{eq:LocRew3}
A_{ij'}(t+1)=0\qquad{}\textnormal{and}\qquad{}A_{ij''}(t+1)=1
\end{equation}
for all possible indexes $j',j''$. The rewiring of all nodes $i$ is performed in parallel. 

These steps are repeated until the condition of Eq.~\ref{eq:StdDev} is fulfilled.

\begin{figure}[ht!]
\begin{center}
$\begin{array}{c}
\epsfxsize=7.5cm \epsffile{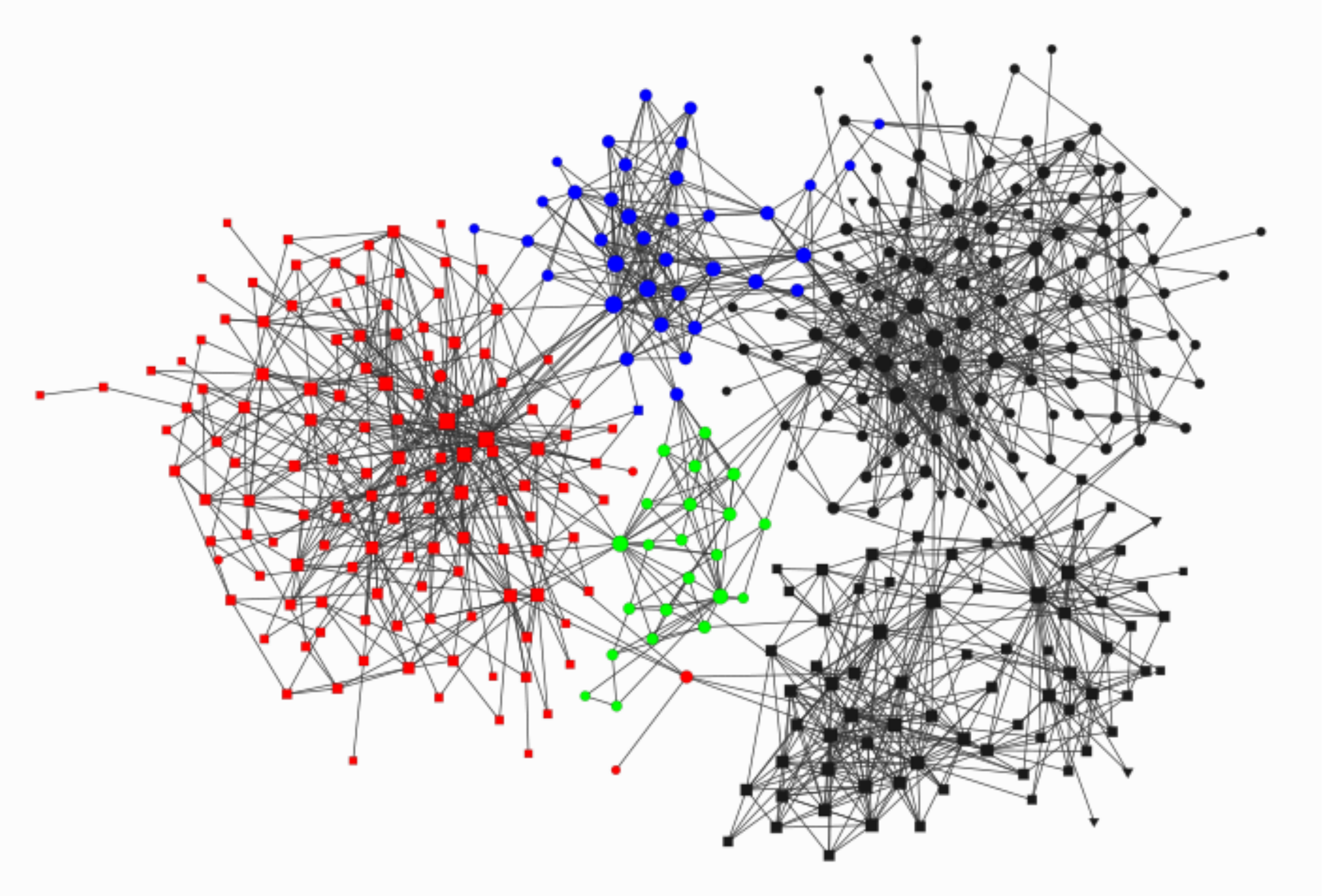}\\
\mbox{\bf (a)}\\
\epsfxsize=7.5cm \epsffile{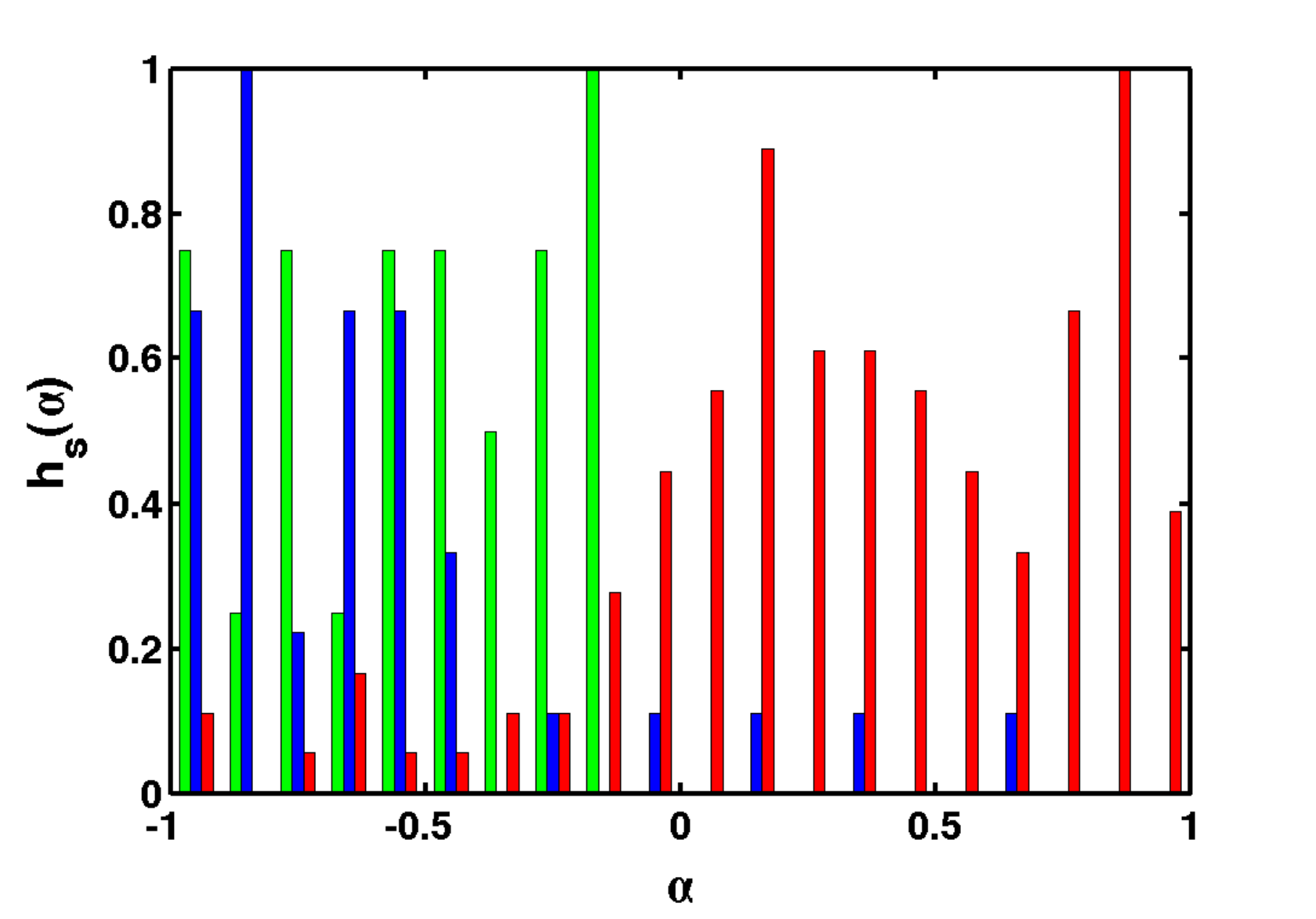}\\
\mbox{\bf (b)}
\end{array}$
\end{center}
\caption{ Results of numerical simulation on a random network of $N=400$ agents keeping $g=1000$, $\langle k_0 \rangle = 4$ and using a Gaussian distribution of opinion and a uniform distribution of $\alpha_i$. (a) A graph of the network with a small community of 23 agents with $x_i=1$ (circles of size proportional to the degree of the node) highlighted in green. A medium size community  of 37 agents in blue, and a large community of 134 agents (squares mean $x=-1$) in red. (b) Histograms normalised by the maximum value of agents in each bin for the three communities highlighted in (a). The colour code is the same as in (a). Observe that the corresponding values of $\alpha$  within the green community are all negative, whilst in the red community most of them are positive. The medium size community shows an intermediate state with most values negative and a few positive.}
 \label{fig:geo}
\end{figure} 

\section{Results}
\label{res}

In Fig.~\ref{fig:geo}(a) we show the results of numerical calculations on a random network of $N=400$ and $\langle k_0\rangle =4$. We observe the formation of at least six detectable communities. The communities can be visually revealed by the graphics software (~\href{http://www.finndiane.fi/software/himmeli/}{himmeli}, developed in our laboratory) because it detects the nodes with more links between them and puts them together. This is not enough to claim the existence of communities. Therefore, we used a fitness algorithm to calculate the optimal partition of the networks~\cite{lancichinetti2009doh}. This algorithm takes the degree of the nodes to group them in well connected clusters, that is, the communities are defined by their high inner degree. We notice in the figure that the opinion of individuals within a cluster is also uniform (circles for positive $x$ and squares for negative $x$), so one can define communities by their common opinion. In the figure we coloured a small community in green, a medium size one in blue and a large community in red.

The interesting new result here is that the parameters $\alpha_i$ for the agents within a topological cluster, are also similar. The initial distribution of $\alpha$ was flat, but the distribution within a given community is not. In Fig,~\ref{fig:geo} (b) we use the same colour code as in (a) to show the histograms made for each of the communities and each one normalized by the maximum value of agents in one bin to keep the height constant. The small communities are formed by agents with negative $\alpha_i$. In communities of medium size the spread of the values of the attitude parameters is larger with few positive ones, and in a large cluster (red) most of them are positive. We then conclude that the negative attitude parameter drives the formation of small communities of individuals with the same attitude and the same opinion (contrary to the majority). In contrast individuals that are agreeing with the overall opinion ($\alpha>0$) tend to group themselves in large communities.

\begin{figure}[ht!]
\begin{center}
\epsfxsize=7.5cm \epsffile{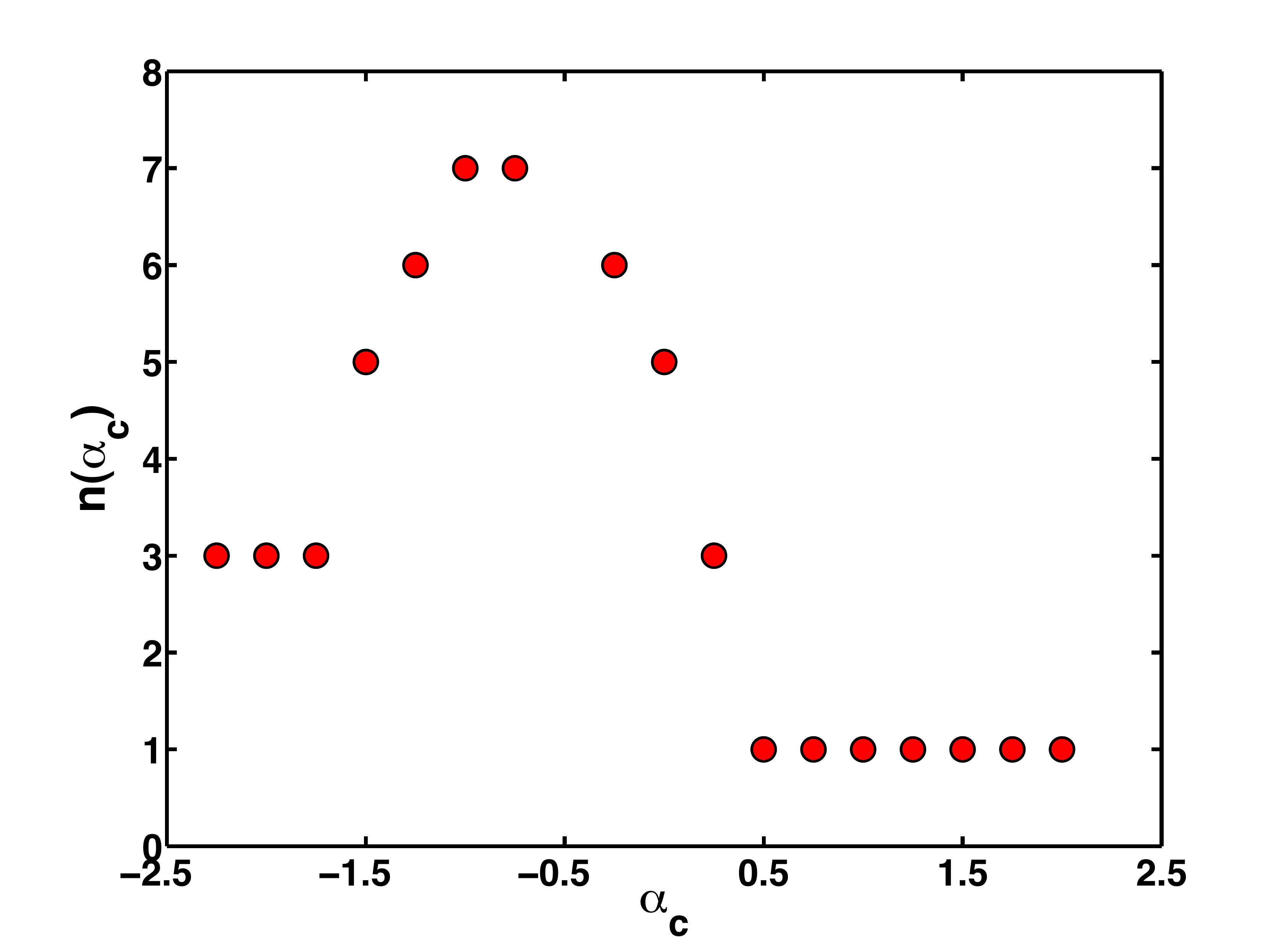}
\end{center}
\caption{Number of detected communities for the final state of a network as a function of the $\alpha$ distribution center. All points correspond to the same realisation of initial opinions over a network with $N = 400$, $\langle k_0 \rangle = 4$, $g = 1000$ and $w = 0.5$. As $\alpha_c$ shifts from large positive to large negative value, the system presents a maximum of 7 in the number of communities between states of consensus ($\alpha_c \geq 0.5$) and 3 detected communities ($\alpha_c \leq -1.75$).}
 \label{fig:centers}
\end{figure}

So far we have considered the case where $\alpha_i$'s are uniformly distributed in the interval [-1,1]. Next we investigate the effect  of picking $\alpha$'s randomly from a narrower and differently positioned distribution on the amount of communities in the system. In Fig.~\ref{fig:centers} we show the number of communities $n(\alpha_c)$ as a function of $\alpha_c$, the center of a uniform $\alpha$ distribution of width $w = 0.5$, for the same realisation of random normally distributed initial opinions, for $N = 400$, $\langle k_0 \rangle = 4$ and $g = 1000$. The communities are detected using the fitness algorithm~\cite{lancichinetti2009doh}. When $\alpha_c \geq 0.5$ (e.g. $\alpha_i \geq 0.25$ for all agents $i$) the system reaches perfect consensus of opinion quite rapidly and only one community is detected. When the $\alpha$'s are allowed to become smaller and even negative, $n(\alpha_c)$ increases until reaching a maximum of 7, a state of the network in which communities can visually be distinguished as well-connected clusters but with few connections between them. When $\alpha_c$ grows more negative the communities start to merge until $n(\alpha_c) = 3$ for $\alpha_c \leq -1.75$. As already mentioned in~\cite{iniguez2009ocf}, the fitness algorithm often reveals a substructure of communities, since for $\alpha_c \leq -1.75$ there are only two communities in terms of opinion sharing.

\section{Concluding remarks}
\label{remarks}
To summarize we have conducted computer simulation studies on our dynamical social network model of opinion formation, which involves fast transaction dynamics between agents through short and long range interactions and slower network rewiring dynamics. These dynamical processes take place with two distinct time scales  but are coupled as $g=T/dt$ being considered as a variable. For the range of intermediate $g$-values the model produces co-evolving opinion and community structure.  In this the key role is played by the uniformly distributed attitude parameter $\alpha$ in the long range term of the transaction dynamics, describing the agent's attitude towards the overall mood of the society.  The network structure was found to consist of small communities of agents whose attitude parameter values are all negative, while for the intermediate size communities there are also a few agents with positive attitude and for large communities the attitude of agents is predominantly positive. Hence we conclude that the attitude parameter of agents serves as the main driving force for opinion and community formation to minimize the amount of overall "frustration" in the system. Interpreting the above in sociological terms would mean that  people with opposing attitude towards the society would feel less frustrated or happier staying in small communities, while people with agreeing attitude are aligning with the society feeling comfortable in a large community of similar people. Carrying this line of thought even further would suggest that according to our model the society would move more dynamically when there is a certain portion of individuals opposing the overall societal opinion.  

 \section*{Acknowledgements}
 G.I. and K.K. acknowledge the Academy of Finland, the Finnish Center of Excellence program 2006 - 2011, under Project No. 129670. K.K. and J.K. acknowledge support from EU's FP7 FET Open STREP Project ICTeCollective No. 238597 and J.K. also support from Grant No. OTKA K60456. K.K. and R.A.B. want to acknowledge financial support from Conacyt through Project No. 79641. R.A.B. is grateful to the Centre of Excellence in Computational Complex Systems Research - COSY of Aalto University for support and hospitality for the visits when most of this work has been done.
 


\bibliographystyle{model1a-num-names}
\bibliography{article}







\end{document}